\documentstyle[aps,epsf,epsfig]{revtex}

\newcommand{\bq}{\begin{equation}}
\newcommand{\eq}{\end{equation}}
\newcommand{\bqn}{\begin{eqnarray}}
\newcommand{\eqn}{\end{eqnarray}}
\newcommand{\nb}{\nonumber}
\newcommand{\lb}{\label}

\begin{document}
\title{Self-Similar Collapse of Scalar Field with Plane Symmetry}
\author{ Anzhong Wang $^{a,\; b}$ \thanks{ E-Mail: Anzhong$\_$Wang@baylor.edu},
Yumei Wu $^{a, \; c}$ \thanks{ E-Mail: yumei@dmm.im.ufrj.br} \and
Zhong Chao Wu $^{a}$ \thanks{ E-Mail: zcwu@zjut.edu.cn }}
\address{ $^{a}$ Department of Physics,
Zhejiang University of Technology, Hang Zhou 310032, P.R. China\\
$^{b}$ CASPER, Physics Department, P.O. Box 97316, Baylor University, Waco, TX76798-7316\\
$^{c}$  Institute of Mathematics,   the Federal University of Rio de Janeiro,  Caixa Postal 68530,
CEP 21945-970, Rio de Janeiro, RJ, Brazil
}

\date{\today }
\maketitle

\begin{abstract}

Plane symmetric self-similar solutions to Einstein's
four-dimensional theory of gravity are studied and  all such solutions are given
analytically in closed form. The local and global properties of these solutions
are investigated  and it is shown that some of the solutions can be interpreted as
representing gravitational collapse of the scalar field. During the collapse,
trapped surfaces are never developed. As a result, no black hole is formed.
Although the collapse always ends with spacetime singularities, it is found
that these singularities are spacelike and not naked.

\end{abstract}

\vspace{.6cm}
\noindent{PACS Numbers: 04.20.Dw  04.20.Jb  04.40.Nr  97.60.Lf }

\section{Introduction}
\lb{SecI}
\renewcommand{\theequation}{1.\arabic{equation}}
\setcounter{equation}{0}

Recently, self-similar solutions to the Einstein field equations have attracted lots of
attention, not only because the corresponding problem can be considerably
simplified and, as a result, can be studied analytically \cite{Car}, but also because
their relevance to critical phenomena in  gravitational collapse, which were first
discovered by Choptuik in 1993 in his numerical study of collapsing spherically
symmetric scalar field  \cite{Chop93}. The phenomena  have lately been found in various matter
fields collapse \cite{Gun00,Wang01}. In particular, one of the present authors
studied analytically the gravitational collapse of cylindrically symmetric scalar field
in four-dimensional spacetimes, and found a class of exact solutions with self-similarity
\cite{Wang03}. It was shown explicitly that one of the solutions has precisely one
unstable mode. By definition, this is a critical solution that sits on the boundary
separating two different phases in the phase space of the initial data. This serves as the
second analytical model of critical collapse found so far. The first analytical
model was found  by Garfinkle in 2001 in $2+1$ gravity \cite{Gar01}, right after the
numerical simulations of Pretorius and
Choptuik \cite{PC00} and  Husain and Olivier \cite{HO01}. The linear perturbations of
Garfinkle's self-similar solutions, denoted by $S[n]$, where $n$ is a parameter, were systematically
studied in \cite{GG02} and \cite{HW02}. Due to different boundary conditions imposed, different results
were obtained. In particular, in \cite{GG02} it was found that the solution with
$n = 3$ has precisely one unstable mode, while in \cite{HW02} it was found that the one with $n = 4$
has one and only one unstable mode. Although the solution with $n =4$ is best matched with the
numerical critical solution found by  Pretorius and  Choptuik, the exponent, $\gamma$, of the
black hole mass
\bq
\lb{1.1}
M_{BH} \propto (p - p^{*})^{\gamma},
\eq
 obtained by \cite{Even}
 \bq
 \lb{1.2}
 \gamma = \frac{1}{\left|k_{1}\right|},
 \eq
is quite different from the numerical one, $\gamma \sim 1.2$ \cite{HW02}, where $k_{1}$ is the
unstable mode of the critical solution. On the other hand, the critical solution found in \cite{GG02}
is different from the numerical one \cite{PC00}, but the resultant exponent, $\gamma = 4/3$, is close
to the numerical one found by  Pretorius and  Choptuik, although it is still quite different from the
numerical one found by Husain and Olivier, $\gamma \sim 0.81$ \cite{HO01}.

In this paper, we have no tendency to resolve the above disputations, but look for more analytical
solutions that might represent critical collapse. In the analytical studies of critical collapse,
the investigation is usually divided into two steps: One first finds some particular solutions by
imposing certain symmetries, such as, homothetic self-similarity (HSS). This can
mathematically simplify the problem considerably. For example, in the spherically symmetric
case, by imposing HSS symmetry the Einstein field equations will be
reduced from PDE's to ODE's.  Once the particular solutions are
known, one can study their linear perturbations and find the
spectra of the corresponding eigen-modes. If a solution has  precisely
one unstable mode,  it may represent a  critical solution,
sitting on a boundary that separates two different basins of attraction in the phase space.
In this paper, we shall restrict ourselves only to the first step. In particular, we
shall  study self-similar solutions of massless scalar field
with plane symmetry in Einstein's four-dimensional theory of gravity, and present
all such solutions in closed form. Then, we shall study their local and global properties, whereby show
some of them can be interpreted as representing gravitational collapse of the scalar field.
During the collapse, trapped surfaces are never developed. As a result, no black hole is formed.
Although the collapse always ends with spacetime singularities, it is found that these singularities
are spacelike and not naked.

\section{Plane symmetric Spacetimes with Self-similarity}
\lb{SecII}
\renewcommand{\theequation}{2.\arabic{equation}}
\setcounter{equation}{0}

The general metric for   spacetimes with plane symmetry   can be
cast in the form \cite{Kramer80},
 \bq
 \lb{2.0}
 ds^{2} = 2 e^{-M(u,v)}du dv  -  e^{-U(u,v)}\left(dx^{2} +
 dy^{2}\right),
 \eq
where $\left\{x^{\mu}\right\} = \{u, v, x, y\}, \; (\mu = 0,1, 2,
3)$, and $- \infty < x^{\mu} < + \infty$. The three Killing
vectors that characterize the symmetry are given by $\xi_{(1)} =
\partial_{x}, \; \xi_{(2)} = \partial_{y}$ and $\xi_{(3)} = x\partial_{y}
- y\partial_{x}$. The non-vanishing components of the Ricci tensor are
given by \cite{TW90}
\bqn
\lb{2.1a}
R_{uu} &=& \frac{1}{2}\left(2U_{,uu} - {U_{,u}}^{2}
+ 2U_{,u}M_{,u}\right),\\
\lb{2.1b}
R_{uv} &=& \frac{1}{2}\left(2M_{,uv} +  2U_{,uv}
- U_{,u} U_{,v}\right),\\
\lb{2.1c}
R_{vv} &=& \frac{1}{2}\left(2U_{,vv} - {U_{,v}}^{2}
+ 2U_{,v}M_{,v}\right),\\
\lb{2.1d} R_{xx} &=& R_{yy} = - e^{M-U}\left(U_{,uv} - U_{,u}
U_{,v}\right), \eqn where $(\;)_{,u} \equiv
\partial(\;)/\partial_{u}$ and so on. A massless scalar field
satisfies the Klein-Gordon equation, $
g^{\alpha\beta}\phi_{;\alpha\beta} = 0$, which in the present case
takes the form, \bq \lb{3.7} 2\phi_{,uv} - \phi_{,u}U_{,v} -
\phi_{,v}U_{,u} = 0. \eq However, this equation is not independent
of the Einstein field equations and can be obtained from the
Bianchi identities $G_{\mu\alpha;\beta}g^{\alpha\beta} = 0$.

Spacetimes with {\em homothetic self-similarity} (or {\em
self-similarity of the first kind}) is usually defined by the
existence of a conform Killing vector $\xi^{\mu}$ that satisfies
the equations \cite{CT71},
 \bq
 \lb{2.2}
 \xi_{\mu;\nu} + \xi_{\nu;\mu} = 2 g_{\mu\nu},
 \eq
where a semicolon ``;" denotes the covariant derivative. It can be
shown \cite{Taub72} that for the spacetimes described by Eq.(\ref{2.0}) there
are two types of self-similar solutions. The first type is given by
 \bqn
 \lb{2.3a}
&& M(u,v) = M(z),\;\;\; U(u, v) = S(z) - 2\ln(-u),\nb\\
&& \xi^{\mu} \partial_{\mu} = u\partial_{u} + v
 \partial_{v},\;\;\;\; z \equiv \frac{v}{(- u)},
 \eqn
while the second type is given by
 \bqn
 \lb{2.3b}
&& M(u,v)= M(z),\;\;\; U(u, v) = U(z),\nb\\
&& \xi^{\mu} \partial_{\mu} = x^{\lambda}\partial_{\lambda},\;\;\;\;
 z \equiv \frac{v}{(- u)}.
 \eqn
Solutions with the properties of Eq.(\ref{2.3a}) will be referred to
as the Type A self-similar solutions, and the ones with the properties of Eq.(\ref{2.3b})
will be referred to as the Type B self-similar solutions. In the following we shall
consider them separately. For the sake of convenience, we shall first
restrict ourselves to the region $u \le 0,\; v \ge 0$. Once we
find the self-similar solutions in this region, we shall extend
them to other regions whenever it is necessary.

\section{ Solutions of Massless Scalar Field with Type A Self-Similarity}
\lb{SecIII}
\renewcommand{\theequation}{3.\arabic{equation}}
\setcounter{equation}{0}

The solutions with Type A self-similarity are characterized by Eq.(\ref{2.3a}),
for which the non-vanishing components of the Ricci tensor
given by Eqs.(\ref{2.1a})-(\ref{2.1d}) become
 \bqn
 \lb{3.4a}
 R_{uu} &=& \frac{z}{u^{2}}\left[z\left(S'' -
 \frac{1}{2}{S'}^{2}\right) + \left(zS' + 2\right)M'\right],\\
 \lb{3.4b}
 R_{uv} &=& \frac{1}{u^{2}}\left[z\left(M'' + S''\right) -
 \frac{1}{2}z{S'}^{2} + M' \right],\\
 \lb{3.4c}
 R_{vv} &=& \frac{1}{u^{2}}\left(S'' - \frac{1}{2}{S'}^{2} +  S'M'\right),\\
 \lb{3.4d}
 R_{xx} &=& R_{yy} = -\frac{1}{u^{2}}e^{M-U}\left[z\left(S'' -
 {S'}^{2}\right) -S'\right],
 \eqn
where a prime denotes the ordinary differentiation with respect to
$z$. For a massless scalar field, the Einstein field equations
read
 \bq
 \lb{3.6}
 R_{\mu\nu} = \kappa \phi_{,\mu}\phi_{,\nu},
 \eq
where $\kappa [\equiv 8\pi G/c^{4}]$ is the  Einstein coupling
constant. In this paper we shall choose units such that $\kappa  =
1$. It can be shown that a massless scalar field
$\phi(u,v)$ that is consistent with spacetimes with homothetic
self-similarity must take the form,
 \bq
 \lb{3.8}
 \phi(u,v) = 2q\ln(-u) + \varphi(z),
 \eq
where $q$ is an arbitrary constant, and $\varphi(z)$ is an
arbitrary function, which will  be determined by the Einstein field
equations (\ref{3.6}). Inserting Eqs.(\ref{3.4a})-(\ref{3.4d}) and
(\ref{3.8}) into Eq.(\ref{3.6}), we find that
 \bqn
 \lb{3.9a}
  z\left[z\left(S'' - \frac{1}{2}{S'}^{2}\right) + \left(zS' + 2\right)M'\right]
 &=& \left(z\varphi' - 2q\right)^{2},\\
 \lb{3.9b}
  z\left(M'' + S''\right) -
 \frac{1}{2}z{S'}^{2} + M' &=& \varphi'\left(z\varphi' - 2q\right),\\
 \lb{3.9c}
  S'' - \frac{1}{2}{S'}^{2} +  S'M' &=& {\varphi'}^{2},\\
 \lb{3.9d}
 \frac{S''}{S'} - S'  - \frac{1}{z} &=& 0.
 \eqn
Eq.(\ref{3.9d}) has the general solution,
 \bq
 \lb{3.10}
 S(z) = - \ln\left|az^{2} + b\right|,
 \eq
where $a$ and $b$ are two integration constants. When $ab = 0$, it
can be shown that the corresponding solutions correspond to null
dust fluid, and in this paper we shall not consider this case.
Thus, from now on we shall assume that $ab \not= 0$. Substituting
the solution (\ref{3.10}) into Eqs.(\ref{3.9a})-(\ref{3.9c}), we
find that
 \bqn
  \lb{3.11a}
   & & M' = - \frac{1}{2az\left(az^{2} + b\right)}\left[2ab
    + \left(az^{2} + b\right)^{2}{\varphi'}^{2}\right],\\
 \lb{3.11b}
    && a\left(z\varphi' - 2q\right)^{2} + b {\varphi'}^{2}
    = - \frac{2ab}{az^{2} + b},\\
 \lb{3.11c}
  && zM'' + M' - \frac{2abz}{\left(az^{2} + b\right)^{2}}
  =  \varphi' \left(z\varphi' - 2q\right).
 \eqn
On the other hand, one can show that Eq.(\ref{3.7}) takes the form
 \bq
 \lb{3.12}
 \varphi'' + \frac{2az}{az^{2} + b} \varphi'
 - \frac{2aq}{az^{2} + b} = 0.
 \eq
It can be shown that Eq.(\ref{3.11c}) is not independent. As a
matter of fact, it can be obtained from Eqs.(\ref{3.11a}) and
(\ref{3.12}). Thus, in the following we need only to consider
Eqs.(\ref{3.11a}), (\ref{3.11b}) and (\ref{3.12}). Integrating
Eq.(\ref{3.12}) we find that
 \bq
 \lb{3.13}
 \varphi'(z) = \frac{2aqz + c_{1}}{az^{2} + b},
 \eq
where $c_{1}$ is a constant. Inserting Eq.(\ref{3.13}) into
Eq.(\ref{3.11a}),  we find that the constant $c_{1}$ must be given
by
 \bq
 \lb{3.14}
 c_{1} = \pm \left[-2ab\left(1 + 2q^{2}\right)\right]^{1/2}.
 \eq
To have $c_{1}$ real, we must require $ab <0$. Then, from
Eqs.(\ref{3.11a}) and (\ref{3.13}) we obtain,
 \bqn
 \lb{3.15}
 M(z) &=& 2q^{2}\ln\left|\frac{z}{az^{2} + b}\right|
     - cq \ln\left|\frac{\alpha - z}{\alpha + z}\right| +
     M_{0},\nb\\
  \varphi(z) &=& q\ln\left|az^{2} + b\right|
     + \frac{1}{2}c \ln\left|\frac{\alpha - z}{\alpha + z}\right| +
     \varphi_{0},
  \eqn
 where $M_{0}$ and $\varphi_{0}$ are other integration constants,
 and
  \bqn
  \lb{3.16}
  c &\equiv& \pm  \left[2\left(1 +
  2q^{2}\right)\right]^{1/2},\nb\\
  \alpha &\equiv& \left(-\frac{b}{a}\right)^{1/2},
  \;\;\; ab < 0.
  \eqn

Rescaling the coordinates $x$ and $y$, without lose of generality,
we can always set $|a| = 1$. Then, the solutions can be finally
written in the form,
 \bqn
 \lb{3.17}
 M(z) &=& 2q^{2}\ln\left|\frac{z}{\alpha^{2} - z^{2}}\right|
     - cq \ln\left|\frac{\alpha - z}{\alpha + z}\right| +
     M_{0},\nb\\
 S(z) &=& - \ln\left|\alpha^{2} - z^{2}\right|,\nb\\
  \varphi(z) &=& q\ln\left|\alpha^{2} - z^{2}\right|
     + \frac{1}{2}c \ln\left|\frac{\alpha - z}{\alpha + z}\right| +
     \varphi_{0}.
  \eqn

To study the above solutions, let us first consider the region
where $\alpha \ge z$. Then, it can be shown that
 \bqn
 \lb{3.18}
 \phi_{,\alpha}  \phi^{,\alpha} &=&
 \frac{4{\alpha}e^{M_{0}}\left(- uv\right)^{2q^{2}}}
 {\left[\alpha(-u)  - v\right]^{2q^{2} + cq + 2}
 \left[\alpha(-u)  + v\right]^{2q^{2} - cq + 2}}\nb\\
& & \times \left[cq\left(v^{2} + \alpha^{2}u^{2}\right)
 - \alpha\left(1+4q^{2}\right) u v\right],\nb\\
 e^{M} &=& \frac{e^{M_{0}}\left(- uv\right)^{2q^{2}}}
 {\left[\alpha(-u)  - v\right]^{2q^{2} + cq}
 \left[\alpha(-u)  + v\right]^{2q^{2} - cq}}, \; (\alpha \ge z).
 \eqn
Since $2q^{2} \pm cq + 2$ is always positive, from the above
expressions we can see that the spacetime is singular on the
hypersurface $v = \alpha (-u)$. The normal vector to this surface
is given by
 \bq
 \lb{3.19}
 n_{\mu} \equiv \frac{\partial\left(v + {\alpha}u\right)}{\partial
 x^{\mu}} = \delta^{v}_{\mu} + \alpha\delta^{u}_{\mu},
 \eq
which is always timelike, as now we have $n_{\alpha}n^{\alpha} = 2
{\alpha}e^{M} > 0$. Therefore, the singularity is spacelike and
services as the up boundary of the region $v \ge 0,\; u \le 0$. On
the other hand, from these expressions we can also see that the
spacetime is free of curvature singularity on the hypersurface $v
= 0$, although the metric coefficient $M$ is singular. Thus, to
have a geodesically maximal spacetime, the solutions need to be
extended beyond this surface. For the sake of convenience, in the
following we shall consider the two cases $2q^{2} < 1$ and $2q^{2}
\ge 1$ separately.

\subsection{$\alpha \ge z,\; 2q^{2} < 1$}

In this case, introducing two new coordinates $\bar{u}$ and
$\bar{v}$ via the relations
 \bq
 \lb{3.20}
 u = - \left(-\bar{u}\right)^{n},\;\;\;
 v = \bar{v}^{n},
 \eq
where
 \bq
 \lb{3.21}
 n \equiv \frac{1}{1 - 2q^{2}},
 \eq
we find that in terms of $\bar{u}$ and $\bar{v}$ the metric takes
the form,
 \bqn
 \lb{3.22}
 ds^{2} &=& \left[\alpha\left(-\bar{u}\right)^{n}
          - \bar{v}^{n}\right]^{2q^{2} + cq}
          \left[\alpha\left(-\bar{u}\right)^{n}
          + \bar{v}^{n}\right]^{2q^{2} - cq}d\bar{u}d\bar{v}\nb\\
          & &
          - \left(\alpha^{2}\bar{u}^{2n} -
          \bar{v}^{2n}\right)\left(dx^{2} + dy^{2}\right),
 \eqn
from which we can see that the metric coefficients are no longer
singular on the hypersurface $v = 0$ or $\bar{v} = 0$. The
coordinate transformations (\ref{3.20}) map the region $v \ge 0,\;
u \le 0$ into the region $\bar{v} \ge 0,\; \bar{u} \le 0$, which
will be referred to as Region $I$, as shown in Fig. 1. Then, the
region $\bar{v} < 0,\; \bar{u} \le 0$ represents an extended
region, which will be referred to as Region $II$. It should be
noted that this extension is not always physically acceptable. In
particular, the extension may not be analytical and the extended
metric coefficients may not be real, unless $n$ is an integer.
When the extension is not analytical, it is also not unique.
Therefore, to have an unique extension, in the following we shall
consider the case where $n$ is an integer. Then, in terms of
$\bar{v}$ and $\bar{u}$ we find that
 \bqn
 \lb{3.23}
 \phi_{,\alpha}  \phi^{,\alpha} &=&
 \frac{4{\alpha}e^{M_{0}}\left(- \bar{u}\bar{v}\right)^{n-1}}
 {\left[\alpha(-\bar{u})^{n}  - \bar{v}^{n}\right]^{2q^{2} + cq + 2}
 \left[\alpha(-\bar{u})^{n}  + \bar{v}^{n}\right]^{2q^{2} - cq + 2}}\nb\\
& & \times \left[cq\left(\bar{v}^{2n} +
\alpha^{2}\bar{u}^{2n}\right)
 + \alpha\left(1+4q^{2}\right)\left(- \bar{u}\bar{v}\right)^{n}\right],
 \eqn
from which we find that $\phi_{,\alpha}$ is timelike only when $n$ is an odd
integer  in the extended region, $II$. The spacetime is always singular on the
hypersurface $\bar{v} = \alpha^{-1/n}\bar{u}$. This singularity is timelike, and the corresponding
Penrose diagram is given by Fig. 1. Due to this singular behavior, it is found difficult
to interpreted the corresponding solutions as representing physically reasonable
model of gravitational collapse.

 \begin{figure}[htbp]
 \begin{center}
 \label{fig1}
 \leavevmode
  \epsfig{file=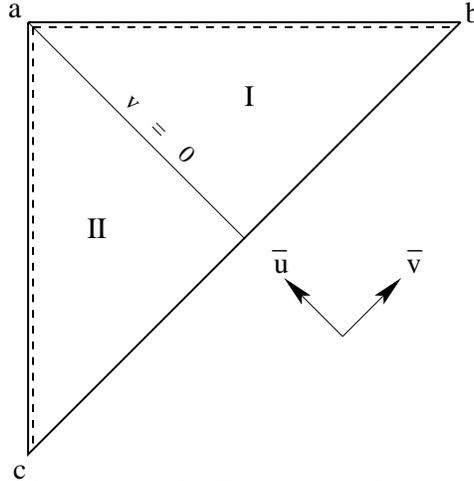,width=0.35\textwidth,angle=0}
 \caption{The Penrose diagram for the solutions given  by
 Eq. (\ref{3.22}) with $n (> 1)$ being an integer. Region $I$ is defined as
 $v \ge 0,\; u \le 0$, and Region $II$ as $v \le 0,\; u \le 0$.
 The  spacetime is singular on the both  horizontal and vertical double
lines $ab$ and $ac$, where $ v = \alpha (-u)$ and $\bar{v} = \alpha^{-1/n}\bar{u}$, 
respectively. The line $bc$ represents the past null infinity $ u = -\infty$.}
 \end{center}
 \end{figure}

\subsection{$\alpha \ge z,\; 2q^{2} \ge 1$}

In this case, the spacetime in the region $v \ge 0, \; u \le 0$ is
already geodesically maximal, and does not need to be extended
beyond the hypersurface $v = 0$. To see this, let us consider the
null geodesics along $u, \; x, \; y = Const.$, which are given by
 \bq
 \lb{3.24}
 \frac{d^{2}v}{d\lambda^{2}}
 - \frac{\partial{M(u_{0},v)}}{\partial{v}}
 \left(\frac{dv}{d\lambda}\right)^{2} = 0,
 \eq
where $\lambda$ denotes the affine parameter along the null
geodesics. Near the hypersurface $v = 0$, it can be shown that the
above equation has the following general solution,
 \bq
 \lb{3.25}
 v(\lambda) = \cases{\left(c_{1}\lambda + c_{2}\right)^{-1/(2q^{2}
 - 1)}, & $2q^{2} > 1$,\cr
 e^{c_{1}\lambda + c_{2}},& $2q^{2} = 1$,\cr}
 \eq
where $c_{1,2}$ are the integration constants. Thus, as $v
\rightarrow 0$, we have $|\lambda| \rightarrow \infty$, that is,
the ``distance" between the point $(u, v) = (u_{0}, 0)$ and any
other one along the null geodesic $u = u_{0}$, say, $(u, v) =
(u_{0}, v_{0})$, where $v_{0} > 0$, is infinitively large. In
other words, the hypersurface $v = 0$ in the present case
represents a physical boundary of the spacetime. Then, the
corresponding Penrose diagram is given by Fig. 2.

 \begin{figure}[htbp]
 \begin{center}
 \label{fig2}
 \leavevmode
  \epsfig{file=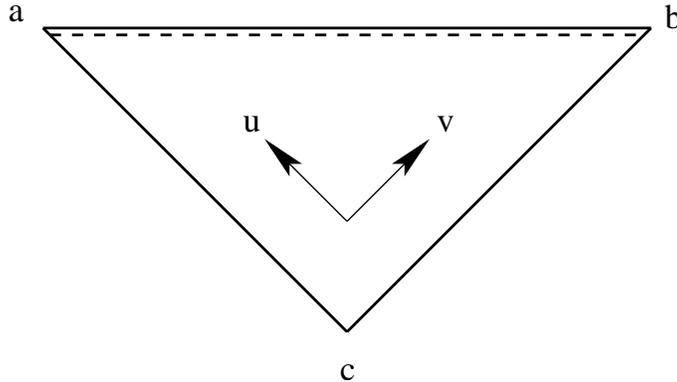,width=0.5\textwidth,angle=0}
 \caption{The Penrose diagram for the solutions given  by
 Eq. (\ref{3.17}) with $\alpha > z,\; 2q^{2} > 1$.
 The  spacetime is singular on  the horizontal double
line $ab$, but not on the ones $ac$ and $bc$ that now serve as the null infinity
boundaries of the spacetime, where the line $ac$ is given by $v = 0$ and the one
$bc$ by $u = - \infty$.}
 \end{center}
 \end{figure}

In the region $u < 0,\; v > 0$, Eq.(\ref{3.18}) shows that the scalar field is timelike.
Thus, one may consider the above solution as representing gravitational collapse of the
scalar field, and the collapse always forms a spacetime singularity on the hypersurface
$v = \alpha (-u)$. Although no trapped surface is formed, the singularity  is not
naked, as it can be seen from Fig. 2.

\section{ Solutions of Massless Scalar Field with Type B Self-Similarity}
\lb{SecIV}
\renewcommand{\theequation}{4.\arabic{equation}}
\setcounter{equation}{0}

For the Type B self-similar solutions, we find that the non-vanishing
components of the Ricci tensor are given by
  \bqn
 \lb{4.2a}
 R_{uu} &=& \frac{z}{u^2} \left[z\left (U^{\prime \prime} -
\frac{1}{2}{U'}^2\right)
+ \left(2 + zM^{\prime}\right)U^\prime \right], \\
 \lb{4.2b}
 R_{uv} &=& \frac{1}{u^2} \left[z\left(M^{\prime \prime} + U^{\prime
\prime} -  \frac{1}{2} U^{\prime 2}\right)
   + M^\prime + U^\prime \right], \\
 \lb{4.2c}
 R_{vv} &=& \frac{1}{u^2} \left(U^{\prime \prime} -
\frac{1}{2}U^{\prime 2} + U^{\prime}M^{\prime} \right), \\
 \lb{4.2d}
R_{xx} &=& R_{yy} = - \frac{z }{u^{2}}e^{M-U}\left(U''  - {U'}^{2} +
\frac{U'}{z}\right).
 \eqn
One can show that the corresponding scalar field has to take the
same form, Eq.(\ref{3.8}). Since $\phi = \phi(u,v)$, we find
that $R_{xx} = \phi_{,x}^{2} = 0$, which together with
Eq.(\ref{4.2d}) yields
 \bq
 \lb{4.3}
 U''  - {U'}^{2} + \frac{U'}{z}= 0.
 \eq
This equation has the general solution
 \bq
 \lb{4.4}
 U = - \ln\left[\alpha\ln(z) + \beta\right],
 \eq
where $\alpha$ and $\beta$ are two integration constants. On the
other hand, it can be shown that  now Eq.(\ref{3.7}) becomes, 
\bq
\lb{4.5} 
\left(z{\varphi'}\right)' + \left(q - z{\varphi'}\right)
U' = 0. 
\eq 
The general solution of it is given by 
\bq 
\lb{4.6}
\varphi(z) = q\ln(z) + {\gamma}\ln\left[\alpha\ln(z) +
\beta\right] + \varphi_{0}, 
\eq 
where ${\gamma}$ and $\varphi_{0}$
are arbitrary constants. Inserting Eqs.(\ref{3.8}), (\ref{4.2c}),
(\ref{4.4}) and (\ref{4.4}) into the Einstein field equations for
the $vv$-component, we find that 
\bq 
\lb{4.7} 
M' =
\frac{1-2q{\gamma}}{z} + \frac{\alpha\left(1-2{\gamma}^{2}\right)}
{2z\left[\alpha\ln(z) + \beta\right]} - \frac{q^{2}}{\alpha
z}\left[\alpha\ln(z) + \beta\right], 
\eq 
which has the  solution
\bq 
\lb{4.8} 
M(z) = \left(1-2q{\gamma}\right)\ln(z) +
\frac{1}{2}\left(1 - 2{\gamma}^{2}\right)\ln \left
\{\left[\alpha\ln(z) + \beta\right]\right\} -
\frac{q^{2}}{2\alpha^{2}}\left[\alpha\ln(z) + \beta\right]^{2} +
M_{0}, 
\eq 
where $M_{0}$ is another integration constant.
Substituting the solutions into the rest of the Einstein field
equations, we find that the constant
  ${\gamma}$ must be given by,
\bq
\lb{4.9}
{\gamma} = \frac{1}{2q}.
\eq
Then, the general solutions   are given by
\bqn
\lb{4.10}
U(z) &=& - \ln\left[\alpha\ln(z) + \beta\right],\nb\\
M(z) &=& -\frac{1}{4q^{2}}\left(1 - 2q^{2}\right)\ln \left\{
\left[\alpha\ln(z) + \beta\right] \right\}
- \frac{q^{2}}{2\alpha^{2}}\left[\alpha\ln(z) + \beta\right]^{2} + M_{0},\nb\\
\varphi(z) &=& q\ln(z) + \frac{1}{2q}\ln\left[\alpha\ln(z) + \beta\right]
+ \varphi_{0}.
\eqn
From the above expressions we can see that to have the solutions valid near
the hypersurface $v = 0$ in the region $u \le 0,\; v \ge 0$, we must assume
\bq
\lb{4.11}
\alpha < 0.
\eq
To study the spacetime near the null hypersurface $v = 0$ further, following the discussions
given in Sec.III.B, one can show that now the hypersurface $v = 0$ also
represents a past null infinity. This null infinity is free of spacetime singularity. This
can be seen, for example, from the Ricci scalar, which now is given by
\bqn
\lb{4.12}
R &=& \phi_{,\alpha}\phi^{,\alpha} =
- \frac{4q^{4}\left[\alpha\ln(z) + \beta\right]^{2} - \alpha^{2}}
{2q^{2}zu^{2} \left[\alpha\ln(z) + \beta\right]^{(1+6q^{2})/4q^{2}}}
\nb\\
&& \times \exp\left\{M_{0} -
\frac{q^{2}}{2\alpha^{2}}\left[\alpha\ln(z) +
\beta\right]^{2}\right\}. 
\eqn 
Thus, $v = 0$ is actually a past
null boundary of the spacetime. On the other hand, from
Eq.(\ref{4.12}) we can see that the spacetime is singular on the
hypersurface 
\bq 
\lb{4.13} 
z = z_{0} \equiv e^{-\beta/\alpha}, 
\eq
which is spacelike and serves as the up boundary of the
spacetime.  The corresponding Penrose diagram is given by Fig. 2.
However, unlike the case discussed in Sec.III.B, now the scalar
field is spacelike near the null hypersurface $v = 0$. Therefore,
the corresponding solution cannot be interpreted as representing
gravitational collapse of the scalar field in the region $u < 0,\;
v > 0$.

\section{Conclusions}

In this paper, we have studied plane symmetric self-similar solutions to Einstein's
four-dimensional theory of gravity and found  all such solutions. We have
also studied the local and global properties of those solutions and found that
some of them can be interpreted as representing gravitational collapse of
the scalar field. During the collapse, trapped surfaces are never developed.
As a result, no black hole is formed. Although the collapse always ends with
spacetime singularities, it is found that these singularities are spacelike and
not naked.

An interesting question is: Do these solutions represent critical collapse?
To answer this question,  one needs to study the linear perturbations of the
solutions and show that there exists one solution that has one and only one
unstable mode. Such a study is out of the scope of this paper, and we wish
to return to this problem in another occasion.

\end{document}